\documentclass[twocolumn,amsmath,amssymb,floatfix,superscriptaddress]{revtex4-1}
\usepackage{bm,graphicx,url,epsf,color}
\usepackage[french]{babel}
\usepackage{wasysym}
\usepackage{MnSymbol}

\begin{document}

\title{Circuit Quantum Simulation of a Tomonaga-Luttinger Liquid with an Impurity}

\author{A. Anthore}
\affiliation{Centre de Nanosciences et de Nanotechnologies (C2N), CNRS, Univ Paris Sud-Universit\'e Paris-Saclay, 91120 Palaiseau, France}
\affiliation{Univ Paris Diderot, Sorbonne Paris Cit\'e, 75013 Paris, France}
\author{Z. Iftikhar}
\affiliation{Centre de Nanosciences et de Nanotechnologies (C2N), CNRS, Univ Paris Sud-Universit\'e Paris-Saclay, 91120 Palaiseau, France}
\author{E. Boulat}
\affiliation{Laboratoire Mat\'eriaux et Ph\'enom\`enes Quantiques (MPQ), Univ Paris Diderot, CNRS, Sorbonne Paris Cit\'e, 75013 Paris, France}
\author{F.D. Parmentier}
\affiliation{Centre de Nanosciences et de Nanotechnologies (C2N), CNRS, Univ Paris Sud-Universit\'e Paris-Saclay, 91120 Palaiseau, France}
\author{A. Cavanna}
\affiliation{Centre de Nanosciences et de Nanotechnologies (C2N), CNRS, Univ Paris Sud-Universit\'e Paris-Saclay, 91120 Palaiseau, France}
\author{A. Ouerghi}
\affiliation{Centre de Nanosciences et de Nanotechnologies (C2N), CNRS, Univ Paris Sud-Universit\'e Paris-Saclay, 91120 Palaiseau, France}
\author{U. Gennser}
\affiliation{Centre de Nanosciences et de Nanotechnologies (C2N), CNRS, Univ Paris Sud-Universit\'e Paris-Saclay, 91120 Palaiseau, France}
\author{F. Pierre}
\email{frederic.pierre@u-psud.fr}
\affiliation{Centre de Nanosciences et de Nanotechnologies (C2N), CNRS, Univ Paris Sud-Universit\'e Paris-Saclay, 91120 Palaiseau, France}

\begin{abstract}
The Tomonaga-Luttinger liquid (TLL) concept is believed to generically describe the strongly-correlated physics of one-dimensional systems at low temperatures.
A hallmark signature in 1D conductors is the quantum phase transition between metallic and insulating states induced by a single impurity.
However, this transition impedes experimental explorations of real-world TLLs.
Furthermore, its theoretical treatment, explaining the universal energy rescaling of the conductance at low temperatures, has so far been achieved exactly only for specific interaction strengths.
Quantum simulation can provide a powerful workaround.
Here, a hybrid metal-semiconductor dissipative quantum circuit is shown to implement the analogue of a TLL of adjustable electronic interactions comprising a single, fully tunable scattering impurity.
Measurements reveal the renormalization group `beta-function' for the conductance that completely determines the TLL universal crossover to an insulating state upon cooling.
Moreover, the characteristic scaling energy locating at a given temperature the position within this conductance renormalization flow is established over nine decades versus circuit parameters, and the out-of-equilibrium regime is explored.
With the quantum simulator quality demonstrated from the precise parameter-free validation of existing and novel TLL predictions, quantum simulation is achieved in a strong sense, by elucidating interaction regimes which resist theoretical solutions.
\end{abstract}

\date{\today}

\maketitle

\section{Introduction}

In condensed-matter physics, a great challenge with abundant technological prospects is to understand the microscopic mechanisms of strongly-correlated phenomena.
However, the complexity of strongly-correlated materials hampers their understanding, even more so since already simplified models often constitute formidable theoretical problems.
Quantum phase transitions, which underpin many such behaviors including high-$T_\mathrm{c}$ superconductivity, may provide a wide-ranging universal framework.
The realization of simple well-characterized systems for the experimental study of the strongly-correlated and quantum critical physics is therefore desirable.
Here the many-body physics at one dimension and a connected quantum phase transition between metallic and insulating states are addressed by means of quantum simulation with a nano-engineered circuit. 

Thirty-five years ago, Richard Feynman pointed out that quantum simulation could provide a powerful workaround for the study of complex quantum systems.
It consists in emulating their physics in a device that is easier to control and measure.
In recent years, realizations of quantum simulators have been demonstrated in a variety of platforms, from neutral atoms and trapped ions to superconducting circuits \cite{Georgescu2014}.
Yet, quantum simulations of physics models that are out of reach of analytical and numerical methods remain wanting.
In the present work, we obtain previously unavailable quantum simulated solutions for the transport across a 1D conductor including a local impurity, which is described by the Tomonaga-Luttinger model.

The Tomonaga-Luttinger model \cite{Luttinger1963,Haldane1981,Giamarchi2003} describes massless 1D electrons in local interactions.
It results in collective `Tomonaga-Luttinger liquid' (TLL) behaviors, which are generally expected for 1D systems at low temperatures whether the interacting particles are bosons, fermions or spins \cite{Haldane1981,Giamarchi2003}.
Experimental observations encompass prominent TLL features (see \cite{Giamarchi2012} for a review): the separation of charge and spin degrees of freedom \cite{Auslaender2002,Jompol2009,Bocquillon2013,Hashisaka2017}, the fractionalization of injected charges \cite{Steinberg2008,Prokudina2014,Kamata2014,Freulon2015}, the emergence of quasiparticles of fractional charge \cite{DePicciotto1997,Saminadayar1997} and first signatures of the quantum phase transition between metallic and insulating states at the ballistic critical point \cite{Chang1996,Bockrath1999,Yao1999,Roddaro2004,Parmentier2011,Mebrahtu2012,Jezouin2013}.
Despite clear signatures of TLL behaviors being observed in a growing number of 1D materials, the challenge remains to achieve a quantitative understanding \cite{Giamarchi2012}.
Notably, the extreme sensitivity of the quantum phase transition to an insulating state, triggered in a TLL by even a single impurity, impedes experimental explorations of real-world 1D conductors.
Important successes were achieved investigating the chiral edge channels in the quantum Hall regime \cite{DePicciotto1997,Saminadayar1997,Roddaro2004,Bocquillon2013,Kamata2014,Freulon2015,Hashisaka2017}, a topologically protected system.
However, complications at fractional filling factors obscure the comparison with TLL predictions \cite{Chang2003,Chklovskii1992,Altimiras2012,Inoue2014}. 
The TLL theory also remains incomplete, in spite of great advances \cite{Giamarchi2003}. 
In particular, the transport across an impurity in a TLL, a revealing probe of the underlying collective physics, still misses a full exact treatment \cite{Kane1992b,Fendley1995,Fendley1995b}.
Furthermore, obtaining quantitative predictions directly from the intrinsic parameters characterizing a physical system constitutes an outstanding challenge.
Here, circuit quantum simulation allows us to bypass some of the experimental and theoretical obstacles, thereby paving the way for investigating a broad range of still elusive TLL and quantum phase transition physics, at a high-precision quantitative level.

The crossover of a 1D conductor comprising a single static impurity toward an insulating state is a trademark TLL signature exposing exotic features \cite{Kane1992b,Pham2000}.
It obeys a universal scaling flow, the determination of which counts among the most theoretically challenging Luttinger physics problems.
Exact analytical solutions of the complete universal conductance flow with respect to voltage and temperature were only obtained for specific intensities of electron-electron interactions, corresponding to Luttinger interaction parameters $K\in\{1/m\}$ ($m\in\mathbb{N}$) \cite{Fendley1995,Fendley1995b}.
Expanding upon these previous works, a novel analytical solution is here obtained for $K=2/3$.
Direct quantitative predictions for a TLL with an impurity also require a connection between the system parameters and the characteristic scaling energy, which determines the location within the universal flow at given voltage and temperature.
This connection is, however, very demanding as it generally involves a full treatment including the passage through the non-universal high-energy regime.
Although power-law dependences for this scaling energy were established in the limits of weak and strong impurities \cite{Kane1992b}, a broader and quantitative understanding for arbitrary scattering remains wanting.
Numerically, a large variety of methods were employed to address the problem of a TLL with an impurity, from Monte Carlo to renormalization group techniques \cite{Moon1993,Leung1995,Meden2002,Enss2005,Vidal2007,Freyn2011,Metzner2012,Lo2014}.
To our knowledge, these methods either address a restricted range of parameters or their exactness is difficult to ascertain mathematically (Appendix~B5).
Very reasonable findings were nevertheless obtained (see e.g. \cite{Moon1993,Leung1995,Enss2005,Freyn2011}), including an agreement with exact analytical results \cite{Kane1992b,Fendley1995,Fendley1995b}.
We believe that it should be possible to cover reliably the full range of parameters by combining different numerical methods.

The paper is organized as follows.
In section~II, we describe the experimental implementation of the TLL model with an impurity for different values of the Luttinger interaction parameter $K\in\{1/2,2/3,3/4,4/5\}$, we present signatures of the quantum phase transition between metallic and insulating states at the ballistic critical point, and we detail the model of our device.
In section~III, we focus on the experimental determination of the universal conductance renormalization flow along the conductor-insulator crossover as temperature is changed.
Then we obtain the quantitative relationship between the characteristic scaling energy and the impurity strength, in section~IV.
The section~V extends our investigation to the out-of-equilibrium regime, at a finite dc bias.
It includes the determination of the different universal scaling curves in the non-equilibrium limit of voltage biases large with respect to the temperature, as well as the non-trivial transition from zero bias.
Finally, we present our conclusions and perspectives in section~VI.
Further technical details and additional measurements are provided in the Appendixes.

\section{Circuit quantum simulator}

The present circuit quantum simulator does not rely on assembling many microscopic individual constituents, a `bottom-up' approach often used e.g. with cold atoms \cite{Georgescu2014}.
Instead, we exploit a direct Hamiltonian mapping between, on the one hand, the Tomonaga-Luttinger model for an infinitely long 1D system of spinless electrons comprising a local scattering center and, on the other hand, a short spin-polarized electronic channel in series with a linear resistance $R=(1/K-1)h/e^2$ [$K=(1+Re^2/h)^{-1}$, $h$ the Planck constant, $e$ the electron charge] \cite{Safi2004,Jezouin2013}.
In essence, as further detailed at the end of this section, the collective TLL excitations can be described as gapless bosonic density modes \cite{Haldane1981} corresponding to the electromagnetic-mode decomposition of a linear resistor in the quantum circuit theory \cite{SCT1992}.
Furthermore, the TLL impurity is straightforwardly implemented by the single-channel electronic contact, of intrinsic scattering strength characterized by the bare (unrenormalized) transmission probability $\tau$ of electrons.

\begin{figure}
\centering\includegraphics[width=86mm]{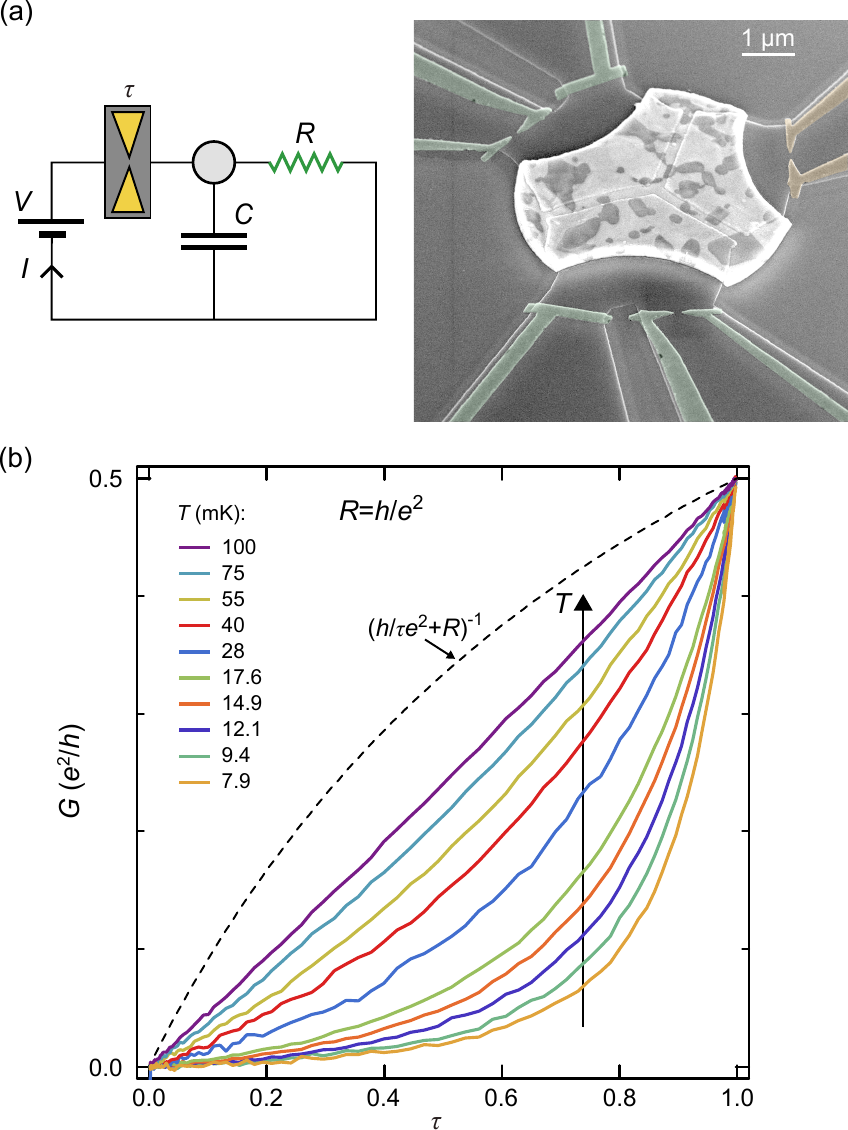}
\caption{
TLL quantum simulator and quantum phase transition between metallic and insulating states.
(a) Device's schematic circuit (left) and colorized $e$-beam micrograph (right).
A QPC set to a single electronic channel of bare transmission probability $\tau$ is formed in a 2DEG with the split gates colorized yellow.
The gates colorized in green control the series resistance $R$ and, thereby, the Luttinger interaction parameter $K=1/(1+Re^2/h)$.
A bright metallic island on top of Y-shaped trenches separates electronic channel and resistance.
(b) Colorized continuous lines display the device's conductance $G=dI/dV$ measured at zero bias ($V=0$) for different temperatures $T$ in the presence of a series resistance adjusted to $R=h/e^2$ ($K=1/2$).
The data are plotted versus bare (unrenormalized) $\tau$, which is determined from $G(V=60\,\mu\mathrm{V},T=7.9\,\mathrm{mK})=(h/\tau e^2+R)^{-1}$ (black dashed line).
The conductance vanishes upon cooling, except at the ballistic quantum critical point $\tau=1$.
}
\normalsize
\label{fig-sample}
\end{figure}

\begin{figure*}
\centering\includegraphics[width=178mm]{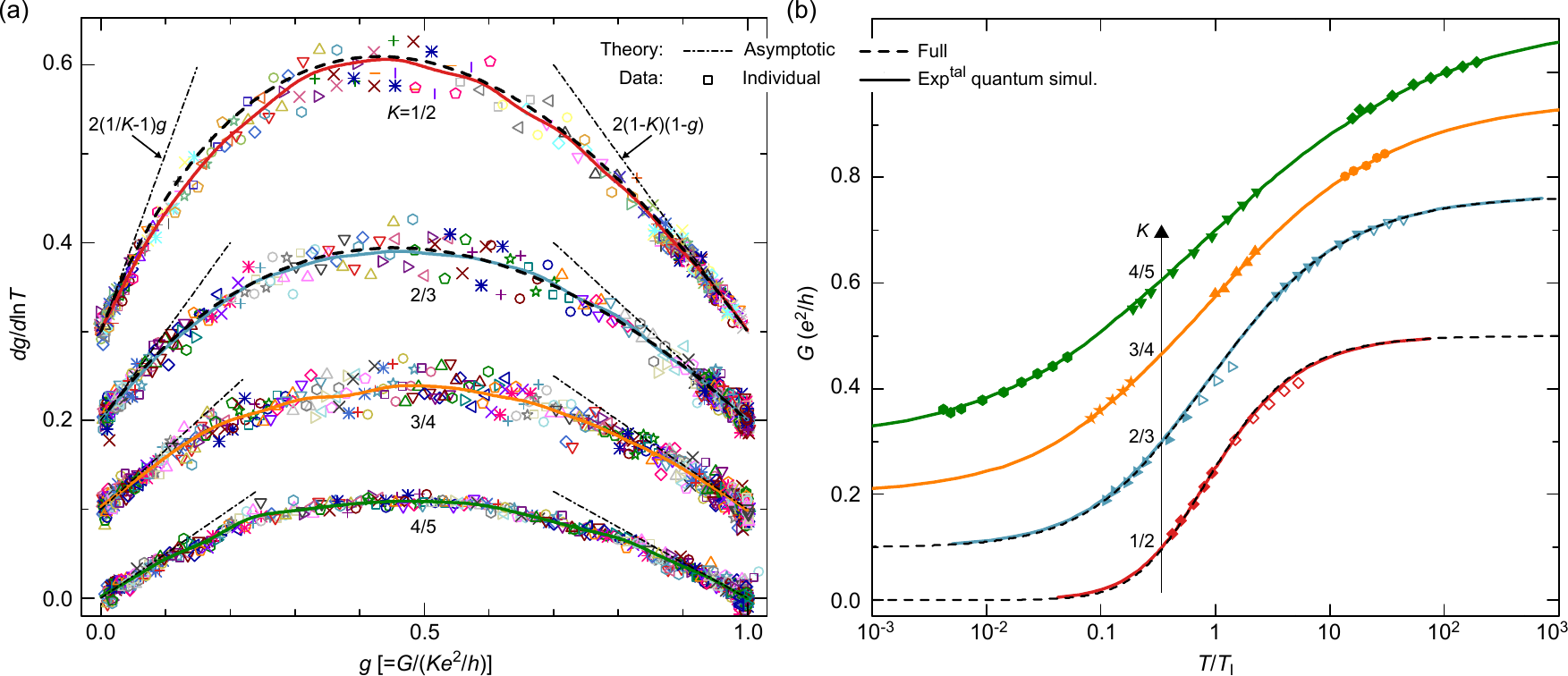}
\caption{
Universal conductor-insulator crossover.
In both panels, data points for a fixed device tuning of $\tau$ at different temperatures are shown with identical symbols.
For clarity, vertical shifts of $0.1$ are applied between $K\in\{1/2,2/3,3/4,4/5\}$ ($R\in\{1,1/2,1/3,1/4\}h/e^2$, respectively).
Colored continuous lines represent the experimental quantum simulated solutions, obtained by averaging the data ensemble.
Exact TLL predictions ($K\in\{1/2,2/3\}$ only) are shown as black dashed lines.
(a) displays the renormalization group beta-functions $\beta_\mathrm{K}(g)$.
Individual data points are the discrete temperature differentiation, at fixed $\tau$, of conductance measurements $\delta g/\delta\ln T$, with $g\equiv G/(Ke^2/h)$.
Black dash-dotted lines show the asymptotic $\beta_\mathrm{K}(g)$ slopes predicted near the $g=0$ and $g=1$ fixed points.
(b) displays the universal conductance flows $G_\mathrm{K}(T/T_\mathrm{I})$.
Quantum simulated curves are obtained by integrating the experimental $\beta_\mathrm{K}(g)$ shown in (a).
Direct conductance measurements are also displayed for representative settings of $\tau$, with $T_\mathrm{I}(\tau,K)$ adjusted at temperatures well-below the high-energy $RC$ cutoff.
Full symbols indicate that the universality criteria $T\lesssim h/25k_\mathrm{B}RC$ is verified.
Non-universal deviations can develop at higher temperatures (open symbols).
}
\label{fig-beta-T}
\end{figure*}

The hybrid metal-semiconductor nanodevice shown in Fig.~1(a) implements such a spin-polarized electronic channel in series with a resistance, as schematically represented.
Expanding upon \cite{Altimiras2007,Parmentier2011,Jezouin2013}, the short electronic channel of fully tunable $\tau\in[0,1]$ is realized by a quantum point contact (QPC) formed in a high-mobility Ga(Al)As two-dimensional electron gas (2DEG) by field effect using split gates.
Other gates are tuned to an adjustable number of ballistic electronic channels $n\in\{1,2,3,4\}$, thereby realizing a linear series resistance $R=h/ne^2$ that corresponds to the Luttinger interaction parameter $K=n/(n+1)$.
Note that a metallic resistance deposited at the surface as in \cite{Parmentier2011} can implement any $K<1$. 
A central metallic island, in essentially perfect electrical contact with the 105\,nm deep 2DEG (see Appendix A), completely breaks the quantum coherence of individual electrons propagating between the QPC and the resistance.
This ascertains that they constitute distinct circuit elements, that are not merged together by the non-local electronic wave functions.
The spin degeneracy is lifted by a perpendicular magnetic field $B=2.7$\,T, corresponding to the quantum Hall regime at filling factor $3$.
All relevant circuits parameters are separately characterized: the series resistance $R=h/ne^2$, the island geometrical capacitance $C\simeq3.1\,$fF that results in the TLL high-energy cutoff $h/RC$ (obtained from Coulomb diamond measurements) and the QPC bare transmission probability $\tau$ (obtained by suppressing the TLL/Coulomb renormalization with a large dc voltage $V\in[45,60]\,\mu$V).
Finally, the experiment relies on a notably precise and reliable determination of the in-situ electronic temperature $T$ down to $7.7$\,mK, at a few-percent accuracy level through shot-noise measurements \cite{Iftikhar2016}.

Figure~1(b) directly illustrates with $R=h/e^2$ ($K=1/2$) the occurrence of a quantum phase transition between metallic and insulating states at the ballistic critical point $\tau=1$.
The zero-bias conductance $G$ across the device is shown versus electronic channel tuning $\tau$ for different temperatures $T$.
At $\tau=1$, $G=1/(h/e^2+R)=Ke^2/h$ does not depend on $T$, which signals a metallic state.
In contrast, any minute back-scattering ($\tau<1$) progressively drives the device away from the unstable metallic fixed point, toward the insulating low-temperature stable fixed point $G=0$ (see Fig.~5 in Appendix for other settings of $R$ and for similar behaviors while varying $V$ instead of $T$).
These data therefore corroborate the $T=0$ quantum phase transition expected for $K<1$ between an insulating state at $\tau<1$ and a conductor at $\tau=1$ (see also the related observations with a resonant level impurity in \cite{Mebrahtu2012,Mebrahtu2013}).

We now provide a more detailed description of the model of our device.
The electronic states transmitted across a single-channel contact can generally be written in terms of adiabatic wave functions having essentially a 1D form.
This form is even more natural in the present integer quantum Hall regime, where spin-polarized electrons propagate along the edges.
One further simplification in our system results from the small Coulomb charging energy $E_\mathrm{C}\simeq0.3\times k_\mathrm{B}$\,K, two to three orders of magnitude smaller than the Fermi and cyclotron energies.
Since TLL physics develops only below the characteristic energy $h/RC=2n\times E_\mathrm{C}$, the spectra of electronic excitations can therefore be linearized with an excellent accuracy, and the unperturbed conduction channel (ballistic limit of the QPC) reads
\begin{equation}
H_\mathrm{0}=i\hbar v_\mathrm{F}\int dx\,\left(\psi^+_{+}\partial_x\psi_{+}-\psi^+_{-}\partial_x\psi_{-}\right),
\end{equation}
with $\psi_{+(-)}$ the annihilation operator for the electrons moving toward (away from) the island and $v_\mathrm{F}$ the Fermi velocity. 
In series with the channel, we engineer a linear impedance $Z(\omega)=R/(1+i\omega RC)$ at $R=h/ne^2$, formed by the $n$ `environmental' ballistic edge channels and the geometrical capacitance of the island.
It can be represented as a Hamiltonian involving an infinite collection of $LC$ oscillators \cite{SCT1992} (or as $n$ ballistic edge channels and a charging energy, see e.g. \cite{Furusaki1995b}).
Physically, it results in Gaussian fluctuations of quantum and thermal origin whose dynamics is dictated by $Z(\omega)$ \cite{SCT1992}:
\begin{align}
\left\langle[\hat{\Phi}(t)-\hat{\Phi}(0)] \hat{\Phi}(0)\right\rangle=\frac{2\hbar^2}{e^2}\int_0^\infty \frac{d\omega}{\omega}\frac{\mathrm{Re}Z(\omega)}{h/e^2}\nonumber \\
\times\left[\coth\left(\frac{\hbar\omega}{2k_\mathrm{B}T}\right)\left[\cos(\omega t)-1\right]-i\sin(\omega t)\right],
\end{align}
with $\hat{\Phi}$ a bosonic operator corresponding to the time integral of the voltage $\hat{u}(t)$ across the impedance ($\partial_t\hat{\Phi}=\hat{u}$).
The coupling between electrons and electromagnetic degrees of freedom simply reads
\begin{equation}
H_\mathrm{C}=-\hat{Q}(V-\partial_t\hat{\Phi}),
\end{equation}
with $\hat{Q}$ the total charge transferred across the QPC (assumed at position $x=0$) given by
\begin{align}
\hat{Q}=&-\frac{e}{2}\left(\int_0^\infty dx\,\left[\psi^+_{+}\psi_{+}+\psi^+_{-}\psi_{-}\right]\right.\nonumber\\
&\left.-\int_{-\infty}^0 dx\,\left[\psi^+_{+}\psi_{+}+\psi^+_{-}\psi_{-}\right]\right).
\end{align}
Finally, the backscattering at the QPC can be modeled as:
\begin{equation}
H_\mathrm{I}=\hbar v_\mathrm{F}r\left[\psi^+_{+}(0)\psi_{-}(0)+\psi^+_{-}(0)\psi_{+}(0)\right],
\end{equation}
with $|r|^2\simeq1-\tau$ for a near ballistic QPC.
Note that the QPC could have also been modeled by a tunneling term between two initially disconnected edge channels.
According to the scattering matrix formalism, these two formulations are essentially equivalent at low energies compared to $E_\mathrm{F}$, as the only relevant QPC parameter is the bare electron transmission probability $\tau$.
We select here the backscattering formulation, which is the more natural choice for fully exploring the crossover from a near ballistic to a disconnected channel.
As previously shown \cite{Safi2004,Jezouin2013}, at sufficiently low energies such that $Z(\omega)\simeq R$, the above model reduces to that of a TLL with a Luttinger interaction parameter $K=1/(1+Re^2/h)$ and a single impurity.
In particular, the present backscattering formulation of the QPC directly matches with the local sine-Gordon model.
Furthermore, beyond the universal low-energy regime, the frequency dependence of $Z(\omega)$ can be seen as a finite-range electron-electron interaction in a 1D conductor (for the specific spatial dependence, see Eq.~S13 in the supplementary information of \cite{Jezouin2013}).

\section{Universal conductor-insulator crossover}

The continuous quantum phase transition theory generally predicts that a system slightly detuned from the critical point follows universal scaling behaviors (along the crossover from `quantum criticality') \cite{Sachdev2011}.
Accordingly, the TLL theory predicts a universal crossover to an insulating state, except at the ballistic quantum critical point $\tau=1$, with all microscopic details encapsulated into an interaction parameter $K$ and a scaling energy $k_\mathrm{B}T_\mathrm{I}$ ($k_\mathrm{B}$ the Boltzmann constant).
As a result, any observable could be recast as a function of $K$, $T/T_\mathrm{I}$ and $eV/k_\mathrm{B}T_\mathrm{I}$.
In our circuit implementation and at zero bias voltage, the conductance reduces to $G(T,R,C,\tau)=G_\mathrm{K}(T/T_\mathrm{I})$.
Such universality is best reformulated into a scale-invariant renormalization flow equation, that does not depend on the convention used to define $T_\mathrm{I}$:
\begin{equation}
\frac{dg}{d\ln T}=\beta_\mathrm{K}(g),
\end{equation}
with $g\equiv Gh/Ke^2$ the dimensionless conductance and $\beta_\mathrm{K}(g)$ the so-called beta-function that fully characterizes the conductance scaling flow.
In Fig.~2, we show theoretical calculations and experimental quantum simulations for both $\beta_\mathrm{K}(g)$ and $G_\mathrm{K}(T/T_\mathrm{I})$ (see below).

Let us first focus on the renormalization group beta-function.
Experimentally, $\beta_\mathrm{K}(g)$ is determined from the discrete differentiation of the measured conductance $\delta g/\delta\ln T$.
In practice, $\delta\ln T$ steps range from $0.3$ to $0.5$ and the temperature is always kept well below the high-energy $RC$ cutoff, with the hottest used temperatures reaching at most $h/25k_\mathrm{B}RC$ ($T\in[8,18]$\,mK with $K\in\{1/2,2/3,3/4\}$, $T\in[40,100]$\,mK with $K=4/5$).
For each $K$, this procedure is repeated at many different QPC tunings (approximately 200 values of $\tau$; identical symbols are used for the same $\tau$).
The pile-up onto the same curve of data points corresponding to different QPC settings (different symbols) provides a direct signature of the predicted universality.
A low-pass Fourier averaging of these data then generates the quantum simulated $\beta_\mathrm{K}(g)$ (fluctuations of small period $\Delta g<0.07$ are filtered out). 
As we now show, the quantitative agreement with available TLL predictions is remarkable, without any adjustable parameter.
Theoretical asymptotic expressions of $\beta_\mathrm{K}(g)$ near the weak and strong impurity limits can be obtained perturbatively, from a poor man's renormalization group approach \cite{Kane1992b}: $\beta_\mathrm{K}(g\ll1)\simeq2(1/K-1)g$, $\beta_\mathrm{K}(1-g\ll1)\simeq2(1-K)(1-g)$.
Beyond these limits, the full exact Bethe ansatz solution of the TLL local (boundary) sine-Gordon model was previously derived for specific values of the interaction parameter $K=1/m$ ($m\in\mathbb{N}$) \cite{Fendley1995,Fendley1995b}.
In addition, we obtain analytically a novel exact solution for $K=2/3$, by developing a different thermodynamic Bethe ansatz (Appendix~B2).
Both the full exact calculations at $K=1/2$ and $2/3$ as well as the asymptotic slopes known for all $K$ are accurately reproduced.
This result firmly establishes our theoretical understanding of the TLL conductor-insulator crossover induced by an impurity as temperature is reduced and, altogether, validates the precise circuit implementation of the TLL-impurity (local sine-Gordon) model.
On these grounds, our measurements reliably provide the quantum simulated beta-function over the complete range $g\in[0,1]$ in the theoretically most challenging regimes $K=3/4$ and $4/5$.

\begin{figure}
\centering\includegraphics[width=86mm]{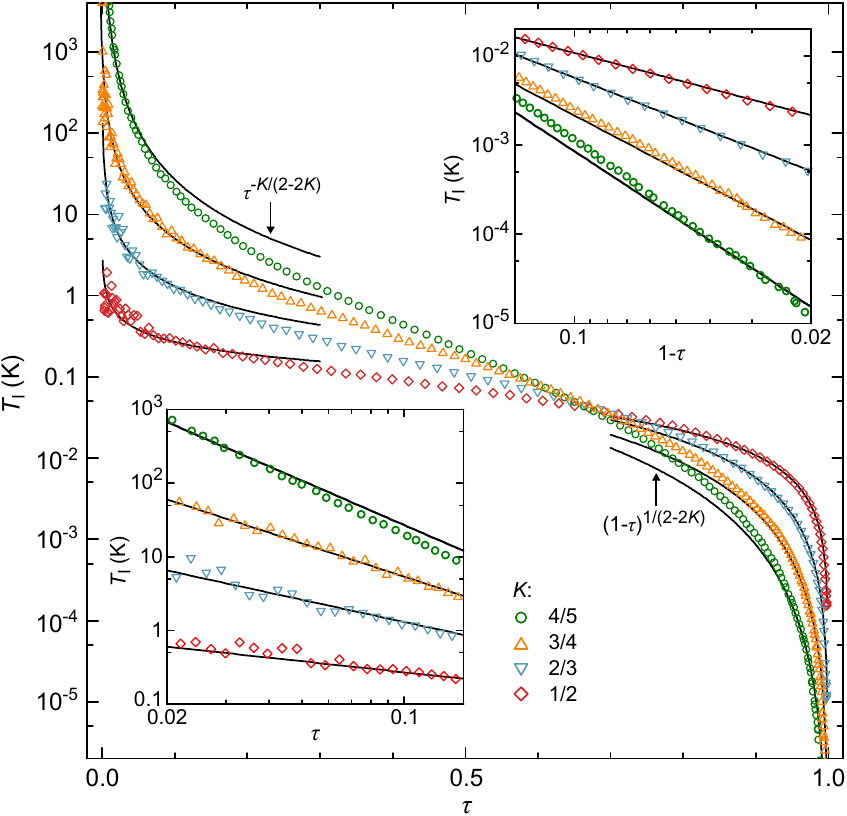}
\caption{
Scaling temperature versus impurity strength. 
Symbols display the experimentally extracted TLL scaling temperature $T_\mathrm{I}$ versus unrenormalized transmission probability $\tau\in[0,1]$.
Each set of identical symbols corresponds to the same tuning of the Luttinger interaction parameter $K$.
Continuous lines represent the asymptotic predictions at $\tau\ll1$ and $1-\tau\ll1$.
The fully quantitative $K=1/2$ prediction is compared to the data without any adjustable parameter.
At other $K\in\{2/3,3/4,4/5\}$, the unknown multiplicative theoretical factor is freely adjusted.
The same data and predictions are also shown in a $\log$-$\log$ scale for $\tau<0.15$ (bottom-left inset) and $\tau>0.85$ (top-right inset).
}
\label{fig-scaling-T}
\end{figure}

The universal renormalization flow $G_\mathrm{K}(T/T_\mathrm{I})$ along the crossover toward an insulating state is then derived from the experimentally quantum simulated $\beta_\mathrm{K}(g)$, by integrating numerically Eq.~1.
Following standard conventions, the scaling TLL temperature $T_\mathrm{I}$ corresponds to a conductance midway between low-$T$ and high-$T$ limits ($G_\mathrm{K}(T=T_\mathrm{I})\equiv(Ke^2/h)/2$).
The obtained experimental quantum simulations of $G_\mathrm{K}(T/T_\mathrm{I})$ span over six orders of magnitude in $T/T_\mathrm{I}$ (Fig.~2(b)).
Comparing without adjustable parameters to the exact calculations at $K=1/2$ and $2/3$ (dashed lines), we find a quantitative agreement on the conductance always better than $0.006e^2/h$ and $0.004e^2/h$, respectively.
Note a previous experimental test of $G_\mathrm{K}(T/T_\mathrm{I})$ at $K=1/5$ (using $T_\mathrm{I}$ as a free parameter and assuming a local fractional quantum Hall filling factor $\nu=1/5$ different from the bulk) \cite{Roddaro2004}.

\section{Scaling temperature versus system parameters}

Within the broad quantum phase transition context, the characteristic crossover energy determines, by delimiting from below, the conditions under which a strongly-correlated `quantum critical' state develops \cite{Sachdev2011}.
In addition, any quantitative prediction for a physical observable at given temperature and voltage requires the knowledge of this reference energy, here corresponding to $k_\mathrm{B}T_\mathrm{I}$, as it provides the dimensionless location within the corresponding universal scaling flow.
However, obtaining the crossover scaling energy directly from the intrinsic system parameters (such as $\tau$, $R$ and $C$) constitutes an important theoretical challenge, usually involving non-universal behaviors at higher energies (see \cite{Altimiras2016} for an alternative theoretical approach of our device in this non-universal regime).
In the vicinity of a continuous quantum phase transition, a power-law increase of the crossover energy with the distance to the quantum critical point is generally expected. 
Its critical exponent, which is identical for any observable, counts among the essential parameters characterizing the transition universality class \cite{Sachdev2011,Vojta2006}.
Regarding the presently investigated TLL-impurity system, a power-law dependence of $T_\mathrm{I}$ is predicted for arbitrary $K$ in both asymptotic limits of weak and strong impurity [$T_\mathrm{I}(1-\tau\ll1)\propto(1-\tau)^{1/(2-2K)}$ and $T_\mathrm{I}(\tau\ll1)\propto\tau^{K/(2-2K)}$] \cite{Kane1992b}.
The circuit TLL quantum simulator allows to test experimentally these predictions and, furthermore, opens access to $T_\mathrm{I}(\tau,R,C)$ over the entire range of $\tau\in[0,1]$.

\begin{figure*}
\centering\includegraphics[width=178mm]{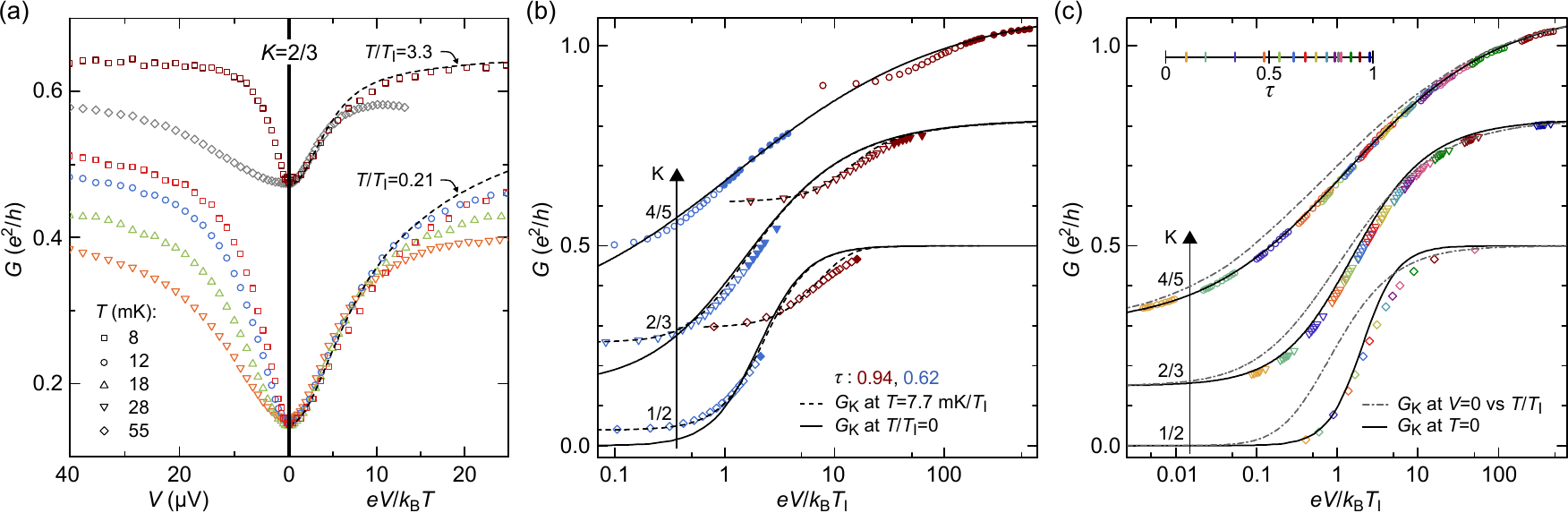}
\caption{
Out-of-equilibrium conductance renormalization. 
Symbols in all panels display the measured differential conductance versus several normalizations of the dc bias voltage $V$. 
For each device setting of $K$ and $\tau$ [color code in (c)], $T_\mathrm{I}$ is separately determined from $G(V=0)$ (Fig.~3).
Theoretical predictions are shown for $T=0$ (continuous lines; see \cite{Fendley1995,Fendley1995b} and Eq.~\ref{FLST=0}) and for values of $T/T_\mathrm{I}$ fixed by the corresponding data (dashed lines; see \cite{Fendley1995,Fendley1995b} and Eqs.~\ref{RK},\ref{currentTBA}).
In (b),(c), we apply $0.15$ vertical shifts between different $K\in\{1/2,2/3,4/5\}$.
(a) illustrates the $eV/k_\mathrm{B}T$ universality.
Data obtained at $K=2/3$ for several settings of $\{\tau,T\}$, each corresponding to the same $G(V=0)$ (two values shown), are plotted versus $V$ (left side) and $eV/k_\mathrm{B}T$ (right side).
(b) shows the non-trivial thermal-nonequilibrium crossover as $V$ is increased, for the representative settings $\tau=0.62$ and $0.94$ at $T=7.7$\,mK.
We show only $|eV|<h/6RC$ data points, sufficiently below the high-energy cutoff.
Large $eV/k_\mathrm{B}T>12$ are signaled by full symbols.
(c) provides a parameter-free comparison between theoretical $G_\mathrm{K}^{T/T_\mathrm{I}=0}(eV/k_\mathrm{B}T_\mathrm{I})$ and data points within $|eV|\in[12k_\mathrm{B}T,h/6RC]$.
The distinct $G_\mathrm{K}^{V/V_\mathrm{I}=0}(T/T_\mathrm{I})$ universal scalings at equilibrium are also plotted, versus $T/T_\mathrm{I}$, as gray dash-dotted lines.
}
\label{fig-ScalingV}
\end{figure*}

The relation between scaling temperature $T_\mathrm{I}$ and circuit parameters $\{\tau,R,C\}$ is obtained experimentally by adjusting to the known $G_\mathrm{K}(T/T_\mathrm{I})$ each set of conductance measurements performed for a fixed device setting ($R$, $\tau$), at temperatures well within the universal TLL regime (the same ranges of $T$ as for $\delta g/\delta\ln T$).
In practice, we compare the data with the theoretically predicted $G_\mathrm{K}(T/T_\mathrm{I})$ when available ($K\in\{1/2,2/3\}$) and with the quantum simulated curves otherwise ($K\in\{3/4,4/5\}$), which all rely on the convention $G_\mathrm{K}(T=T_\mathrm{I})\equiv(Ke^2/h)/2$.
Figure~2(b) illustrates the precision of this procedure at representative settings of $\tau$, with full symbols pointing out $T\lesssim h/25k_\mathrm{B}RC$.
Note for future reference that, although non-universal deviations to $G_\mathrm{K}$ can develop at $T>h/25k_\mathrm{B}RC$ [open symbols in Fig.~2(b)], they remain relatively small up to our maximum temperature of 100\,mK (approximately $h/6k_\mathrm{B}RC$ for $K=1/2$; at $K=3/4$, the device is measured only up to 18\,mK).
The experimentally extracted $T_\mathrm{I}$ span over nine orders of magnitude while varying $\tau$ from 0 to 1 (Fig.~3).
The agreement with the theory at $\tau\lesssim0.15$ and $\tau\gtrsim0.85$ establishes the predicted asymptotic power-laws (see the insets for a $\log$-$\log$ scale comparison).
For $K=1/2$, the asymptotic predictions also include the numerical value of the proportionality coefficient (Appendix~B4) \cite{SCT1992,Furusaki1995b}, which is here experimentally validated without any adjustable parameter.
Note that such quantitative agreement at $K=1/2$ further demonstrates the precise quantum simulation of the circuit's model described section II, including at high temperatures where it does not reduce to the local sine-Gordon model since the capacitance $C$ cannot be neglected.
Beyond these asymptotic limits, our experimental quantum simulations provide specific quantitative predictions at intermediate impurity strengths.

\section{Out-of-equilibrium regime}

The investigation is now extended to out-of-equilibrium situations, by applying a dc bias voltage $V$.
The conductance scaling curves can be markedly different in the non-equilibrium limit ($T=0$) compared to those at equilibrium ($V=0$).
However, the quantum phase transition theory generically predicts the same characteristic energy $k_\mathrm{B}T_\mathrm{I}$.
For a non-zero voltage and temperature, the conductance can therefore be recast as a universal function of $eV/k_\mathrm{B}T_\mathrm{I}$ and $T/T_\mathrm{I}$ or, equivalently, of $eV/k_\mathrm{B}T$ and $T/T_\mathrm{I}$.
Here, we experimentally establish the uniqueness of the scaling energy $k_\mathrm{B}T_\mathrm{I}$ and the distinct shapes of the non-equilibrium conductance curves, as well as the previous and novel theoretical predictions for the TLL model with an impurity at several interaction strengths $K$. 

Typical conductance measurements at $K=2/3$ are plotted versus voltage bias in the left-hand side in Fig.~4(a).
The displayed settings of $\tau$ [color code shown in Fig.~4(c)] and the temperature $T\in[8,55]$\,mK are chosen to have a matching conductance at $V=0$ [two distinct values of $G(V=0)$ shown] and, therefore, correspond to the same $T/T_\mathrm{I}$.
The expected singular (unique) character of the scaling energy $k_\mathrm{B}T_\mathrm{I}$, applying to both voltage and temperature, then translates into an identical dependence as a function of $eV/k_\mathrm{B}T$.  
This uniqueness is directly checked by plotting the same data versus $eV/k_\mathrm{B}T$ in the right-hand side in Fig.~4(a).
We observe for sufficiently low voltages with respect to the high-energy cutoff ($eV\ll h/RC\approx0.1$\,meV), the collapse on a single curve of data sets measured at broadly varying temperatures, different from each other by a factor of up to seven.
The dashed lines represent the theoretical predictions without any additional parameter than the value of $T/T_\mathrm{I}$ fixed by the corresponding zero-bias conductance.
The crossover from thermal ($eV\ll k_\mathrm{B}T$) to non-equilibrium ($eV\gg k_\mathrm{B}T$) regimes is further investigated in Fig.~4(b), for different $K\in\{1/2,2/3,4/5\}$.
At a finite temperature, the full thermal to non-equilibrium crossover upon increasing $V$ is exactly calculated at $K=1/2$ \cite{Fendley1995,Fendley1995b} and here at $K=2/3$ (Appendix~B2).
We systematically observe a good agreement between the conductance data at $eV<h/6RC$ and quantitative crossover predictions, including the expected crossings with $G_\mathrm{K}^{T/T_\mathrm{I}=0}(eV/k_\mathrm{B}T_\mathrm{I})$.
For relatively large bias voltages, the finite temperature calculations converge toward $G_\mathrm{K}^{T/T_\mathrm{I}=0}$, with discrepancies systematically smaller than $0.01e^2/h$ at $eV/k_\mathrm{B}T\gtrsim12$.
In Fig.~4(c), we specifically study the universal conductance scaling in the non-equilibrium limit $G_\mathrm{K}^{T/T_\mathrm{I}=0}$, which is known exactly for all $K$ in contrast to finite temperatures \cite{Fendley1995,Fendley1995b}.
Marked differences with the corresponding temperature ($V=0$) scaling $G_\mathrm{K}(T/T_\mathrm{I})$ are expected to develop as interactions get stronger (as $K$ is reduced).
This difference is illustrated by the distinct shapes of the continuous and dash-dotted lines.
Only those data points expected close to this limit [$eV\in[12k_\mathrm{B}T,h/6RC]$, shown as full symbols in Fig.~4(b)] are displayed. 
Note that relatively small but significant deviations between data and theory appear at $K=2/3$ and further develop at $K=1/2$.
We attribute these to a heating of the central metallic island by the injected Joule power at a finite bias voltage.
Electron-phonon cooling is indeed very inefficient, while outgoing electronic heat flow is lower with a larger series resistance and further reduced by heat Coulomb blockade \cite{Sivre2018}.
The device, therefore, does not truly operate as a quantum simulator in the non-equilibrium regime, as such heating is not included in the TLL model.
Nevertheless, the overall agreement with the theory remains remarkable, especially since it is a direct comparison without any adjustable parameters.
Moreover, heating could, in principle, be taken into account in a refined but fully characterized model involving the known electron-phonon heat flow (Appendix~A1).

\section{Conclusion and outlook}

We have realized an analog quantum simulator for the Tomonaga-Luttinger model with a single impurity, using a broadly tunable and fully characterized quantum circuit.
The device was operated at four different values of the Luttinger interaction parameter ($K\in\{1/2,2/3,3/4,4/5\}$), for which we completely determined the distinct conductance scaling flows to an insulating state as well as the relations connecting the scaling energy to the impurity scattering strength and also explored the non-trivial crossovers from thermal to non-equilibrium regimes.
For $K\in\{1/2,2/3\}$, the quantitative match between data and theory, without any fitting parameter, establishes experimentally the yet untested TLL predictions, including our novel $K=2/3$ exact analytical solution. 
As the observed precise agreement also demonstrates the device's quantum simulator quality, the investigation of interaction regimes that remain both analytically and numerically challenging provides novel quantum simulated solutions for the full conductance scaling flow (for $K\in\{3/4,4/5\}$) and for the characteristic TLL scaling energy (at intermediate scattering strengths).
In practice, it is now possible to evidence 1D correlated physics by comparing with these quantum simulations. 
Beyond Luttinger liquids, the significance of our results extends to the general field of continuous quantum phase transitions, whose study in simple and well-controlled nano-engineered circuits was still limited to universality classes connected with the Kondo effect \cite{Potok2007,Mebrahtu2012,Mebrahtu2013,Jezouin2013,Keller2015,Iftikhar2015,Iftikhar2018}.
Notably, our data establish in a different context the generic expectations of universal scaling behaviors and of a parameter-space power-law broadening for quantum criticality upon increasing the temperature.
Finally, the device's quantum point contact of fully adjustable bare transmission probability $\tau$ emulates an arbitrary single-channel short coherent conductor.
This work therefore directly addresses the modified transport properties of quantum components when embedded into a circuit, a Coulomb induced quantum phenomenon.
The present investigation goes beyond the known dynamical Coulomb blockade limit of a tunnel junction ($\tau\ll1$) in series with a linear impedance \cite{SCT1992}.
Compared to previous experiments at intermediate $\tau$ \cite{Altimiras2007,Parmentier2011,Jezouin2013}, we here broadly investigate the universal regime arising at low temperatures.

The demonstrated TLL quantum simulator opens the path to in-depth quantitative investigations of various facets of correlated physics.
These encompass statistical, thermal and dynamical phenomena now accessible within the present circuit implementation, and include natural implications for quantum nanoelectronic engineering \cite{Parmentier2011,Altimiras2014,Sivre2018}.
Notably, such hybrid metal-semiconductor circuits provide a coveted gateway for exploring novel exotic quasiparticles (in particular the fractional TLL quasiparticles of anyonic statistics \cite{Pham2000}, further than the restricted fractions and avoiding the obscuring complications of quantum Hall systems), the quench dynamics of quantum phase transitions (by driving the TLL transition between metallic and insulating states through a rapid variation of the local impurity or of the global Luttinger parameter), or the 1D correlated physics beyond the short-range interaction paradigm of the TLL model (by the circuit nano-engineering of finite-range electron-electron interactions \cite{Jezouin2013}).

\begin{acknowledgements}
This work was supported by the French RENATECH network and the national French program `Investissements d'Avenir' (Labex NanoSaclay, ANR-10-LABX-0035).

F.P. performed the experiment with inputs from Z.I.;
A.A. and F.P. analyzed the data;
F.D.P. fabricated the sample with inputs from A.A.;
A.C., A.O., and U.G. grew the 2DEG;
E.B. developed the theoretical calculations ($K=2/3$);
F.P. led the project and wrote the manuscript with A.A. and input from E.B. and U.G.
\end{acknowledgements}
 
\section*{Appendix A: Experimental Methods} 

\subsection*{1. Sample}
The sample consists of a Ga(Al)As two-dimensional electron gas buried $105$~nm below the surface (density $2.5\,10^{11}\,\mathrm{cm}^{-2}$, mobility $10^6\,\mathrm{cm}^2\mathrm{V}^{-1}\mathrm{s}^{-1}$). 
Its nanostructuration is performed by standard $e$-beam lithography, dry etching, and metallic deposition. 
The central metallic island (nickel [30\,nm], gold [120\,nm] and germanium [60\,nm]) forms an ohmic contact with the 2DEG (by thermal annealing at 440\,$^\circ$C for 50\,s).
The quality of this ohmic contact is completely characterized, through the individual determination of the electron reflection probability at the metal-2DEG interface for each connected quantum Hall channel.
We find a negligible reflection probability, below $\lesssim0.001\%$ (the statistical uncertainty) for all used channels.
The typical electronic level spacing in the metallic island is estimated to be smaller by more than four orders of magnitude than the thermal energy ($\delta\approx k_\mathrm{B}\times0.2\,\mu$K). 
Combined with the low number of outgoing channels (up to five), this small spacing completely ascertains that the quantum coherence of individual electrons is broken between the QPC and the series resistance.
The sample is tuned in the integer quantum Hall regime at filling factor three. 
This tuning not only breaks the spin degeneracy of electronic channels, but also allows for perfectly transmitted (ballistic) channels across the QPCs, thanks to the topological protection of the chiral quantum Hall channels.
Finally, the charging energy of the island $E_\mathrm{C}\equiv e^2/2C\simeq k_\mathrm{B}\times 0.3$\,K is obtained by standard Coulomb diamond characterization, from the dc voltage height $V_\mathrm{diam}$ of the observed diamonds ($E_\mathrm{C}=eV_\mathrm{diam}/2$).
Note that, in the non-equilibrium regime, an improved device modeling including the island's Joule heating by the applied dc voltage bias could be developed without unknown parameters, using the electron-phonon heat transfer previously obtained for this metallic island: $J_\mathrm{heat}^\mathrm{ph}\simeq3.9\,10^{-8}(T_\Omega^{5.85}-T^{5.85})$\,W, with $T_\Omega$ the electrons' temperature in the island \cite{Sivre2018}.
Note also that the same sample was recently used in \cite{Iftikhar2018}.
The most important difference with this previous work is that the device was tuned in a regime where the island's charge is quantized in \cite{Iftikhar2018}, which allowed us to implement the `charge' equivalent of a magnetic Kondo impurity.
In contrast, there is no trace of charge quantization here, because at least one connected channel is set in the ballistic regime \cite{Matveev1995,Iftikhar2016}.

\subsection*{2. Experimental setup} 
The device is fixed to the mixing chamber plate of a cryofree dilution refrigerator.
Electrical measurement lines connected to the sample include several filters and thermalization stages, as well as two shields at the base temperature that screen spurious high-frequency radiations (see \cite{Iftikhar2016} for further details on the same setup).
Conductances are measured by standard lock-in techniques at low frequencies, below 200~Hz.

\begin{figure*}
\centering\includegraphics [width=140mm]{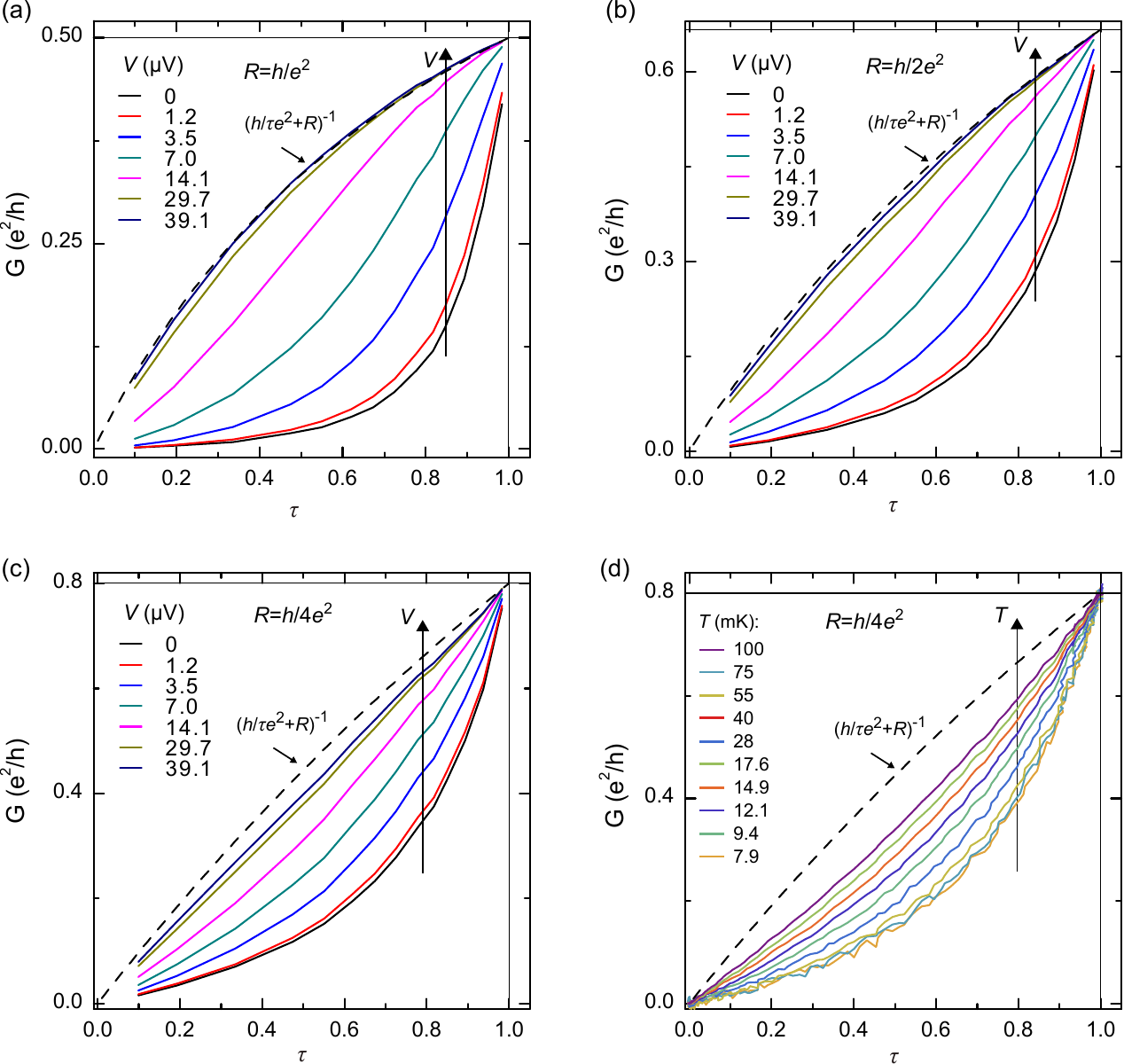}
\caption{
Quantum phase transition between metallic and insulating states.
(a,b,c) Colorized continuous lines display the device's conductance $G=dI/dV$ measured at $T=7.7$~mK and different dc bias voltage $V$ for a series resistance set to $R=\{1,1/2,1/4\}\times h/e^2$ ($K=\{1/2,2/3,4/5\}$).
The data are plotted versus bare transmission probability $\tau$.
(d) Colorized continuous lines display $G=dI/dV$ measured at $V=0$ for different $T$ and a series resistance set to $R=h/4e^2$ ($K=4/5$) versus bare transmission probability $\tau$.
Whatever $K$ ($R$), the conductance progressively vanishes as $V$ or $T$ is reduced except at the ballistic quantum critical point $\tau=1$.
}
\end{figure*}

\subsection*{3. Results' reproducibility}

Figure 5 provides additional evidence of the quantum phase transition between metallic and insulating states at the ballistic critical point, both at different settings of the Luttinger interaction parameter and also from the voltage bias dependence.

The robustness of our results on the conductance scaling flow and on the characteristic scaling energy are further ascertained by implementing an equivalent circuit configuration with a different physical realization of the QPC for the series resistances $R\in\{1,1/2\}\times h/e^2$: The QPC is additionally formed with the top-left split gate colorized green in Fig.~1(a) (instead of the yellow split gate in the article).
As shown in Fig.~6(a), the experimental beta-functions (continuous lines for the data shown in the main article and dash-dotted lines for the additional data) extracted in both equivalent configurations are very close, although the noise level is larger for the additional data.
These beta-functions lead to indistinguishable $G_\mathrm{K}(T/T_\mathrm{I})$ [continuous and dash-dotted lines in the inset in Fig.~6(b)].
The relationships between $T_\mathrm{I}$ and $\tau$ are also identical at experimental accuracy, as shown in the main panel in Fig.~6(b).

\begin{figure*}
\centering\includegraphics [width=140mm]{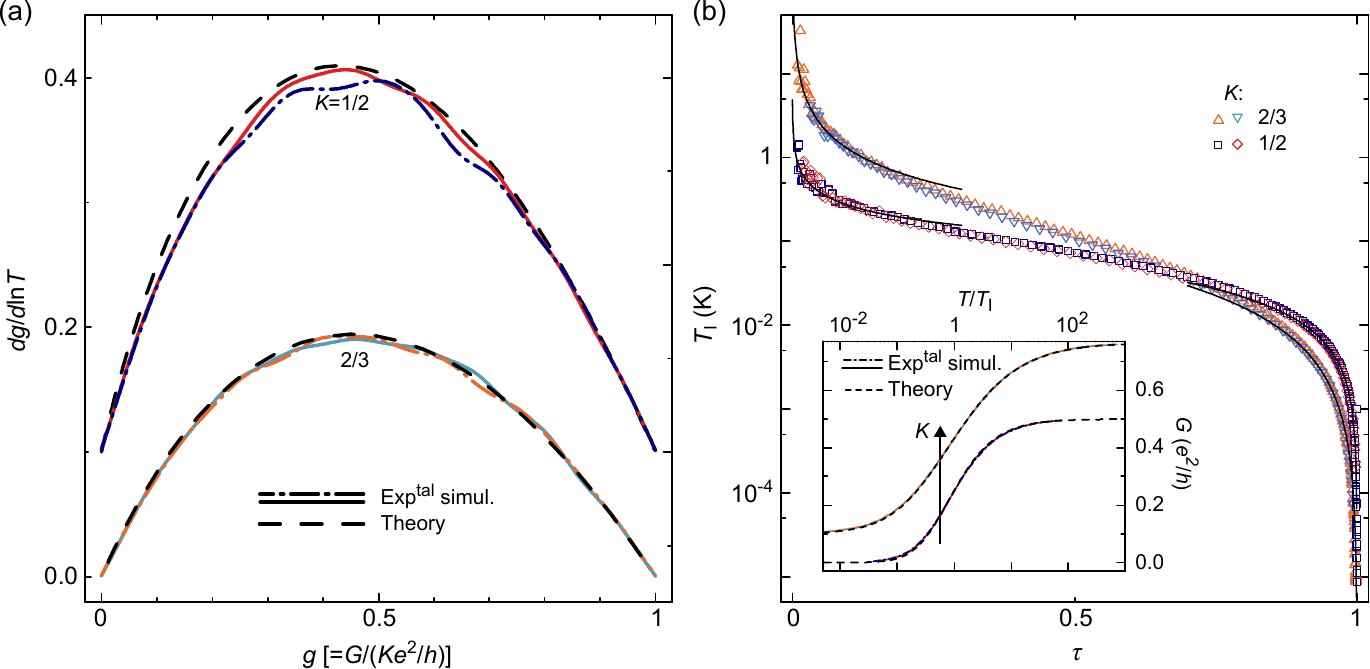}
\caption{
Reproducibility in equivalent circuit configurations.
(a) Renormalization beta-function for two physical implementations of the QPC in series with a linear resistance $R\in\{1,1/2\}\times h/e^2$.
The continuous lines show the same experimental quantum simulations displayed Fig.~2(a), while the dash-dotted lines are obtained with another QPC (at a lower signal-to-noise level).
(b) The experimental scaling temperature $T_\mathrm{I}$ (symbols) does not depend on the impurity realization by different QPCs at our experimental accuracy (different superimposed symbols correspond to different physical realizations of the QPC).
Inset : Equivalence of the simulated renormalization flows $G_\mathrm{K}(T/T_\mathrm{I})$ (continuous lines, main article data; dash-dotted lines, additional data).
}
\end{figure*}

\section*{Appendix B: Theoretical model and predictions}

\subsection*{1. Full scaling flow predictions for $K=1/2$}

The TLL-single impurity model (the local sine-Gordon model) is exactly solved for $K=1/m$ ($m\in\mathbb{N}$) versus arbitrary combinations of the voltage and temperature \cite{Fendley1995,Fendley1995b}.
For $K=1/2$ (corresponding to $R=h/e^2$), the prediction for the universal conductance renormalization curve reduces to a simple analytical expression:
\begin{align}
G&_\mathrm{K}^{T/T_\mathrm{I}}(eV/k_\mathrm{B}T_\mathrm{I})&\nonumber\\
&=\frac{e^2}{2h}\left\{1-\frac{c_1 T_\mathrm{I}}{2\pi T}\mathrm{Re}\left[\Psi'\left(\frac{1}{2}+\frac{c_1 T_\mathrm{I}}{2\pi T}+\frac{ieV}{4 \pi k_\mathrm{B} T}\right)\right]\right\}, \label{RK}
\end{align}
where $\Psi$ is the digamma function.
The numerical factor $c_1\simeq1.01$ is adjusted to fulfill the conventional criterion $G_\mathrm{K}(T=T_\mathrm{I})=(Ke^2/h)/2=e^2/4h$ (at $V=0$).
Note that we use here the expression given in the arXiv version of \cite{Fendley1995b}, which corrects the expression given in the Physical Review papers \cite{Fendley1995,Fendley1995b} by a factor of $1/2$ on the voltage bias dependence.
With this correction, Eq.~\ref{RK} precisely matches asymptotically, at $\max[eV,k_\mathrm{B}T]\ll k_\mathrm{B}T_\mathrm{I}$, with both the well-tested dynamical Coulomb blockade theory \cite{SCT1992,Odintsov1991,Iftikhar2016} and a new perturbative Keldysh calculation (see section B4).
Equation~\ref{RK} quantitatively corresponds to the $K=1/2$ theory displayed in Figs.~2 and 4.
Remarkably, the same renormalization function applies to the conductance along the crossover from two-channel Kondo quantum criticality to a Fermi liquid in the charge Kondo implementation \cite{Furusaki1995b,Mitchell2016,Iftikhar2018}.

\subsection*{2. Novel full scaling flow predictions for $K=2/3$}

We develop a different thermodynamic Bethe ansatz to account for the full voltage and temperature dependence of the universal conductance renormalization curve at $K=2/3$ ($R=h/2e^2$). 
The tunneling in a resistive environment problem maps onto the non-equilibrium local (boundary) sine-Gordon model \cite{Safi2004} (at sine-Gordon coupling $\beta=\sqrt{8\pi K}$), which has been solved exactly \cite{Fendley1995,Fendley1995b} when $K=1/m$ with $m\in\mathbb{N}$. 
The solution relies on the integrability of the sine-Gordon model and proceeds by identifying the quasiparticles (the sine-Gordon solitons $S^+$ and antisolitons $S^-$), diagonalizing the boundary interaction, and then averaging the current operator in a thermal gas of solitons and antisolitons, which are interacting particles (they scatter among each other non trivially).
However, when $K^{-1}\notin\mathbb{N}$, the scattering between quasiparticles is no longer diagonal; i.e., the scattering process $(S^+(p),S^-(p'))\longrightarrow (S^-(p),S^+(p'))$, with $p$ the momentum of solitons, has a non-vanishing amplitude. 
Since the solitons can change their internal quantum number during a scattering event, it is no longer possible to derive the thermodynamics of the gas of quasiparticles. 

This complication is overcome by building explicitly \emph{new} quasiparticle modes by means of the algebraic Bethe ansatz \cite{Sklyanin1980}:
Those modes are technically obtained as states diagonalizing the transfer matrix.
For $K=2/3$, this procedure yields quasiparticles $A_a$ ($a\in\{s,0,\bar0\})$ diagonalizing the  scattering between quasiparticles \cite{Fendley1992} and making it possible to do the thermodynamics of this new gas.
In this new gas of interacting quasiparticles, $A_s$ is a neutral object that describes indifferently solitons or antisolitons and carries the kinetic energy, whereas $A_{0,\bar0}$ are massless quasiparticles (no kinetic energy) with charge $\pm2e$.
A (positive) voltage bias tends to populate $A_0$ states and to deplete $A_{\bar0}$ states and is accounted for by an appropriate chemical potential on the $A_{0,{\bar0}}$ quasiparticles.

By making use of the boundary Yang-Baxter equation, we then can derive the boundary scattering matrix that mixes the modes $A_0$ and $A_{\bar 0}$.
We thus have at hand all the ingredients for a complete and exact description of the system out of equilibrium:
(i) a gas of interacting quasiparticles with known thermodynamics (even including the voltage bias) and (ii) a boundary scattering matrix describing electrical transfers across the junction.

The net result is that the electrical current through the structure can be exactly written as
\begin{equation}
I(V,T,\tilde T_{\mathrm{I}})=\frac{2 q_s k_\mathrm{B} T}{h}\int\mathrm{d}\theta \;\left[\rho_0(\theta)-\rho_{\bar{0}}(\theta)\right]\mathcal{T}_\mathrm{I}(\theta)
\end{equation}
where $q_s=e$ is the electrical charge of the sine-Gordon solitons, $\theta=\ln(v_\mathrm{F}p/k_\mathrm{B}T)$ is a rapidity (the logarithm of the energy of an individual particle) parametrizing the momentum $p$ of solitons ($v_\mathrm{F}$ is the Fermi velocity in the 2DEG), and ${\cal T}_\mathrm{I}=(1+e^{-2(\theta-\ln\tilde T_{\mathrm{I}}/T)})^{-1}$ is the probability that an incoming $A_0$ quasiparticle be scattered as an outgoing $A_{\bar0}$ particle at the junction. 
Note that, in this approach, the scaling temperature $\tilde T_\mathrm{I}$ (we refer to it as the `TBA scale') is defined by the relation ${\cal T}_\mathrm{I}(p=k_\mathrm{B}\tilde T_\mathrm{I}/v_\mathrm{F})=\frac 12$, i.e. an incoming  quasiparticle of type $A_0$  is scattered as an outgoing $A_{\bar0}$ quasiparticle with probability $1/2$.
The  densities of quasiparticles  $\rho_a(\theta)$ are obtained via a thermodynamical Bethe ansatz 
on the gas of interacting quasiparticles of type $A_a$, and are written in a standard way using the so-called pseudo-energies $\epsilon_a(\theta)$. 
The current in terms of the pseudo-energies reads
\begin{align}
I{\textstyle \left( \frac T{\tilde T_{\mathrm{I}}},  \frac {eV} {k_\mathrm{B}\tilde T_{\mathrm{I}}} \right)}=&\frac{2e k_\mathrm{B}T}{h}\!\! \int_{-\infty}^\infty
\frac{\mathrm{d}\theta}{1+\left(\frac {\tilde T_{\mathrm{I}}}T e^{-\theta}\right)^2}\nonumber\\
&\times \partial_\theta\ln\frac{1+e^{-\epsilon_0(\theta)+\frac {eV}{k_\mathrm{B}T}}}{1+e^{-\epsilon_0(\theta)-\frac {eV}{k_\mathrm{B}T}}},
\label{currentTBA}
\end{align}
whereas the $\epsilon_a$ are determined by integral equations:
\begin{align}
\epsilon_0=& -\frac{1}{2\pi}\frac{1}{\cosh\theta}\star \ln(1+e^{-\epsilon_s}),\\
\epsilon_s=& e^\theta -\frac{1}{2\pi}\frac{1}{\cosh\theta}\star \ln\left[(1+e^{-\epsilon_0+\frac {eV}{k_\mathrm{B}T}})\right.\nonumber\\
&\left.\times(1+e^{-\epsilon_0-\frac {eV}{k_\mathrm{B}T}})\right].
\end{align}
This prediction quantitatively corresponds to the $K=2/3$ theory displayed in Figs.~2 and 4.
Based on this full solution, we find $c_1=\tilde T_\mathrm{I}/T_\mathrm{I}\simeq0.77$ for the proportionality coefficient between the TBA scale $\tilde T_\mathrm{I}$ in Eq.~\ref{currentTBA} and the scaling TLL temperature $T_\mathrm{I}$ used in the manuscript [which is defined using the standard convention $G_\mathrm{K}(T=T_\mathrm{I})=(Ke^2/h)/2=2e^2/6h$, at $V=0$].

\subsection*{3. Non-equilibrium ($T=0$) scaling predictions for arbitrary $K$}
In a first step, we write down the $T=0$ predictions in terms of the TBA temperature scale $\tilde T_\mathrm{I}$ used in \cite{Fendley1995,Fendley1995b} and also used to formulate the new $K=2/3$ prediction in the previous section (Eq.~\ref{currentTBA}).
Then, we connect $\tilde T_\mathrm{I}$ with the scaling TLL temperature $T_\mathrm{I}$ defined such as $G_\mathrm{K}(T=T_\mathrm{I})\equiv(Ke^2/h)/2$ at $V=0$.
These predictions are used for the non-equilibrium theory curves shown in Fig.~4.

First, the differential conductance $G\equiv dI/dV$ is calculated from the derivative of the current written at $T=0$ in terms of two different power series for the regimes of high and low voltages, which together cover the full range of voltages \cite{Fendley1995,Fendley1995b}:
\begin{equation}\label{FLST=0}
  I(V)=\left\{
  \begin{array}{l}
    \displaystyle \frac{VKe^2}{h}\left[1-K\sum_{n=1}^\infty a_n(K)\times\left(\frac{V}{V_\mathrm{I}}\right)^{2n(K-1)}\right] \\ \\
    \displaystyle \frac{e^2V}{h}\sum_{n=1}^\infty a_n(\frac{1}{K})\times\left(\frac{V}{V_\mathrm{I}}\right)^{2n\left(\frac{1}{K}-1\right)}\; ,
  \end{array}
  \right.
\end{equation}
where the functions $a_n(x)$ read
\begin{equation}
  a_n(x)=(-1)^{n+1}\frac{\sqrt{\pi}\,\Gamma(n x)}{2\Gamma(n)\Gamma\left[\frac{3}{2}+n(x-1)\right]},
\end{equation}
and with the scaling voltage $V_\mathrm{I}$ related to the TBA temperature scale $\tilde T_\mathrm{I}$ through \cite{Fendley1995b}
\begin{equation}\label{VBvsTBtilde}
eV_\mathrm{I} = \frac{2\sqrt{\pi}\Gamma(\frac1{2(1-K)})}{K\;\Gamma(\frac K{2(1-K)})}k_\mathrm{B}\tilde T_\mathrm{I}.
\end{equation}

Second, we determine the quantitative factor $c_1\equiv\tilde T_\mathrm{I}/T_\mathrm{I}$ connecting the TBA temperature scale (and, therefore, $V_\mathrm{I}$) with the scaling TLL temperature.
It is most straightforward if the conductance is known for arbitrary $T/\tilde T_\mathrm{I}$ at $V=0$.
In that case, $c_1$ is simply given by the temperature ratio $\tilde T_\mathrm{I}/T$ for which the conductance takes the value $(Ke^2/h)/2$, since it also corresponds to $T=T_\mathrm{I}$.
In practice, we find $c_1\simeq1.01$ and $0.77$ for $K=1/2$ and $K=2/3$, respectively.
In the absence of a full theoretical solution versus $T/\tilde T_\mathrm{I}$, such as for $K=4/5$ in Fig.~4, a different approach is needed as now detailed. 
In essence, the low-temperature and low-voltage asymptotic functions for the conductance are connected one to another through new perturbative calculations (an identical connection can also be made using the dynamical Coulomb blockade theory).
This link establishes a bridge between the quantum simulated equilibrium conductance curve $G_\mathrm{K}(T/T_\mathrm{I})$ and the non-equilibrium ($T=0$) predictions of Eq.~\ref{FLST=0} versus $eV/k_\mathrm{B}\tilde T_\mathrm{I}$, thereby connecting $T_\mathrm{I}$ and $\tilde T_\mathrm{I}$. 
Now, more specifically, we establish asymptotic results in the limit $\max[eV,k_\mathrm{B}T]\ll k_\mathrm{B}T_\mathrm{I}$ by means of a Keldysh perturbative calculation of the current, yielding (E.\;Boulat \textit{et al.}, in preparation)
\begin{equation}
I( T,V) =   A_\mathrm{K} \left(\frac T{T_\mathrm{I}}\right)^{\frac2K-1}\;\mathrm{Im}\left[ \frac{\Gamma(K^{-1}+i\frac {eV}{2\pi k_\mathrm{B} T})}{\Gamma(1-K^{-1}+i\frac {eV}{2\pi k_\mathrm{B} T})}  \right]
\label{IKPT}
\end{equation}
where $A_\mathrm{K}$ is a numerical prefactor not needed here.
Expression \ref{IKPT} leads to the prediction of a universal ratio between the conductance in the low-voltage $T=0$ regime, on the one hand, and the low-temperature $V=0$ regime, on the other hand:
\begin{align}
\frac{G(T=0,eV/k_\mathrm{B}\ll T_\mathrm{I})}{G(T\ll T_\mathrm{I},V=0)} &=\frac{2K^{-1}-1}{\Gamma(K^{-1})^2}\; \left(\frac{eV}{2\pi k_\mathrm{B} T}\right)^{\frac{2}{K}-2}
\nonumber\\
&=\alpha_\mathrm{KPT}(K)\times\Big(\frac{eV}{k_\mathrm{B}T}\Big)^{\frac{2}{K}-2}.
\label{alphaKPT}
\end{align}
Note that the same ratio can be obtained from the asymptotic, low-energy limit predictions of the dynamical Coulomb blockade theory using the corresponding series resistance $R=(K^{-1}-1)\frac{h}{e^2}$ \cite{SCT1992,Odintsov1991,Iftikhar2016}.
On the one hand, the low-voltage conductance asymptote is given by the first term of the lower series in Eq.~\ref{FLST=0}, and involves as the only unknown variable the scaling TBA temperature $\tilde T_\mathrm{I}$.
On the other hand, the low-temperature conductance asymptote at zero bias voltage scales as
\begin{equation}
G_\mathrm{K}(T\rightarrow0,V=0)=\frac{e^2}{h}b(K)\times\left(\frac{T}{T_\mathrm{I}}\right)^{\frac{2}{K}-2},
\label{GtllTto0}
\end{equation}
with $b(K)$ a numerical coefficient that depends on the quantitative definition of the scaling temperature $T_\mathrm{I}$, here based on the standard convention $G_\mathrm{K}(T=T_\mathrm{I})\equiv(Ke^2/h)/2$.
At $K=4/5$, in the absence of a full theoretical prediction at equilibrium, this convention is implemented using the experimentally quantum simulated solution, which gives $b(4/5)\simeq0.97$, and, comparing the resulting low-voltage to low-temperature ratio with Eq.~\ref{alphaKPT}, we obtain $c_1\equiv\tilde T_\mathrm{I}/T_\mathrm{I}\simeq0.58$.

\subsection*{4. Exact quantitative predictions for $T_\mathrm{I}$ versus physical parameters at $K=1/2$}

Although the asymptotic powerlaw behaviors $T_\mathrm{I}(\tau\ll1)\propto\tau^{\frac{-K}{2-2K}}$ and $T_\mathrm{I}(1-\tau\ll1)\propto(1-\tau)^{\frac{1}{2-2K}}$ are known for arbitrary values of $K$ \cite{Kane1992b}, the numerical prefactor is generally unknown and depends on the specific, non-universal `high-temperature' physics.
In particular, since the studied circuit reduces to the local sine-Gordon model only at low energies with respect to the capacitive $h/RC$ cutoff, the local sine-Gordon solutions obtained at $K=1/m$ ($m\in\mathbb{N}$) and here at $K=2/3$ are not sufficient to connect quantitatively $T_\mathrm{I}$ to the circuit parameters $\tau$, $R$ and $C$.
However, in the special case $K=1/2$ and for the present circuit implementation ($R=h/e^2$), exact quantitative results have previously been obtained for the conductance versus physical parameters \cite{Furusaki1995b,Matveev2002}, thereby giving access to the numerical value of $T_\mathrm{I}$ at $\tau\ll1$ and $1-\tau\ll1$ as detailed below.

First, for $\tau\ll1$, the conductance reads at asymptotically low temperatures and zero bias voltage (Eq.~34 in \cite{Matveev2002} with $r=0$ and $G_L=\tau e^2/h$; note that the exact same prediction can be obtained from the dynamical Coulomb blockade theory \cite{SCT1992})
\begin{equation}
G_\mathrm{MA}(T\rightarrow0)=\tau\frac{e^2}{h}\frac{2 \pi^4}{3 \exp(2\gamma) E_\mathrm{C}^2}(k_\mathrm{B}T)^2.
\end{equation}
This prediction can be matched with the first term of the $T/T_\mathrm{I}\rightarrow0$ series expansion of the TLL $K=1/2$ analytical expression Eq.~\ref{RK}:
\begin{equation}
G_\mathrm{K=1/2}(T\rightarrow0)=\frac{e^2}{h}\frac{\pi^2}{6}\Big(\frac{T}{c_1T_\mathrm{I}}\Big)^2,
\label{FLS_r=1_DLT}
\end{equation}
with $c_1\simeq1.01$.
By identification, we find
\begin{equation}
k_\mathrm{B}T_\mathrm{I}^{\tau\ll1}=\frac{\exp(\gamma)E_\mathrm{C}}{2\pi c_1\sqrt{\tau}},
\end{equation}
which is displayed without adjustable parameters for $\tau<0.3$ as the $K=1/2$ (lower left) black continuous line in Fig.~3.

Second, for $1-\tau\ll1$, the conductance reads at asymptotically low temperatures and zero bias voltage (Eqs.~38 and 26 in \cite{Furusaki1995b})
\begin{equation}
G_\mathrm{FM}(T\rightarrow0)=\frac{e^2}{2h}\left(1-\frac{\exp(\gamma) E_\mathrm{C}}{2\pi T}(1-\tau)\right).
\end{equation}
The corresponding first term in the series expansion of the analytical prediction Eq.~\ref{RK} at $T/T_\mathrm{I}\rightarrow+\infty $ reads
\begin{equation}
G_\mathrm{K=1/2}(T\rightarrow0)=\frac{e^2}{2h}(1-\frac{\pi c_1T_\mathrm{I}}{4T}).
\end{equation}
By identification, we find
\begin{equation}
k_\mathrm{B}T_\mathrm{I}^{1-\tau\ll1}=\frac{2\exp(\gamma)E_\mathrm{C}}{\pi^2c_1}(1-\tau),
\end{equation}
which is displayed without adjustable parameters for $\tau>0.7$ as the $K=1/2$ (higher right) black continuous line in Fig.~3.\\

\subsection*{5. Numerical investigations of 1D systems with an impurity}
The main challenge with exact 1D lattice simulations, including density-matrix renormalization group approaches \cite{Schollwock2005}, is that capturing the physics down to low energies would require an extremely large number of sites (about $10^4$-$10^5$; see \cite{Meden2002}).
This can be achieved within approximate schemes such as the truncated functional renormalization group method that is most reliable for weak interactions (small $1-K$) \cite{Meden2002,Enss2005,Metzner2012} or the multiscale entanglement renormalization ansatz that best captures the low-energy behaviors \cite{Vidal2007,Lo2014}.
Beside lattice simulations, the local sine-Gordon model for infinitely long 1D systems in the universal TLL regime with one impurity can be addressed exactly by quantum Monte Carlo or numerical renormalization group methods.
However, as exact quantum Monte Carlo solutions are usually computed only at discrete imaginary times \cite{Moon1993}, obtaining the dc conductance involves an analytical continuation whose outcome depends on the critical choice of the functional used to fit the numerical data \cite{Giamarchi2003,Leung1995}.
This difficulty can be avoided in real-time quantum Monte Carlo implementations \cite{Leung1995} at the cost of reintroducing the so-called `dynamical sign problem', which limits how low in energy the computations can be made.
A particularly powerful approach to the local sine-Gordon model \cite{Freyn2011} is provided by a more recently developed numerical renormalization group treatment, which allows for the unambiguous determination of the conductance.
Most of the parameters' range is accessible by this approach, although it is numerically challenging to address the limit case of small interaction strengths ($1-K\ll1$).
Note that, to our knowledge, there is at this time no numerical simulation of the full circuit model quantum simulated by our device and described in section II (including deviations from the local sine-Gordon model at large energies where the capacitance is not negligible).
Such full model simulations are required to connect system parameters (here $\tau$, $R$, and $C$) and the scaling energy ($T_\mathrm{I}$)


\end{document}